\documentclass[sigconf,natbib=true]{acmart}
\usepackage{amsmath,nccmath,tabularx}
\usepackage{algorithm}
\usepackage[noend]{algorithmic}
\usepackage{color}
\usepackage{graphicx}
\graphicspath{{images/}}
\usepackage{wrapfig,lipsum}
\usepackage{tabularx}
\usepackage{graphicx}
\usepackage{lscape}
\usepackage{subfigure}
\usepackage{latexsym}
\usepackage{pifont}
\usepackage{multirow}

\usepackage{enumitem}
\setlist{leftmargin=3mm}
\usepackage[T1]{fontenc}
\usepackage[utf8]{inputenc}
\usepackage[font=small,labelfont=bf]{caption}

\newcommand{\SUB}[1]{\ENSURE \hspace{-0.15in} \textbf{#1}}

\AtBeginDocument{%
	\providecommand\BibTeX{{%
			\normalfont B\kern-0.5em{\scshape i\kern-0.25em b}\kern-0.8em\TeX}}}

\copyrightyear{2023} 
\acmYear{2023} 
\setcopyright{acmlicensed}\acmConference[ICTIR '23]{Proceedings of the 2023 ACM SIGIR International Conference on the Theory of Information Retrieval}{July 23, 2023}{Taipei, Taiwan}
\acmBooktitle{Proceedings of the 2023 ACM SIGIR International Conference on the Theory of Information Retrieval (ICTIR '23), July 23, 2023, Taipei, Taiwan}
\acmPrice{15.00}
\acmDOI{10.1145/3578337.3605117}
\acmISBN{979-8-4007-0073-6/23/07}

\begin{document}

\title{An Analysis of Untargeted Poisoning Attack and Defense Methods for Federated Online Learning to Rank Systems}


\author{Shuyi Wang}
\affiliation{
	\institution{The University of Queensland}
	\streetaddress{4072 St Lucia}
	\city{Brisbane}
	\state{QLD}
	\country{Australia}}
\email{shuyi.wang@uq.edu.au}

\author{Guido Zuccon}
\affiliation{
	\institution{The University of Queensland}
	\streetaddress{4072 St Lucia}
	\city{Brisbane}
	\state{QLD}
	\country{Australia}}
\email{g.zuccon@uq.edu.au}


\begin{abstract}
Federated online learning to rank (FOLTR) aims to preserve user privacy by not sharing their searchable data and search interactions, while guaranteeing high search effectiveness, especially in contexts where individual users have scarce training data and interactions. For this, FOLTR trains learning to rank models in an online manner -- i.e. by exploiting users' interactions with the search systems (queries, clicks), rather than labels -- and federatively -- i.e. by not aggregating interaction data in a central server for training purposes, but by training instances of a model on each user device on their own private data, and then sharing the model updates, not the data, across a set of users that have formed the federation. Existing FOLTR methods build upon advances in federated learning.

While federated learning methods have been shown effective at training machine learning models in a distributed way without the need of data sharing, they can be susceptible to attacks that target either the system's security or its overall effectiveness. 

In this paper, we consider attacks on FOLTR systems that aim to compromise their search effectiveness. Within this scope, we experiment with and analyse data and model poisoning attack methods to showcase their impact on FOLTR search effectiveness. We also explore the effectiveness of defense methods designed to counteract attacks on FOLTR systems. We contribute an understanding of the effect of attack and defense methods for FOLTR systems, as well as identifying the key factors influencing their effectiveness.

\end{abstract}

\ccsdesc[500]{Information systems~Learning to rank}

\keywords{Online Learning to Rank, Federated Learning, Federated Online Learning to Rank, Model Poisoning, Data Poisoning}

\maketitle

\section{Introduction}

\begin{figure}[t]
	\centering
	\includegraphics[width=1\columnwidth]{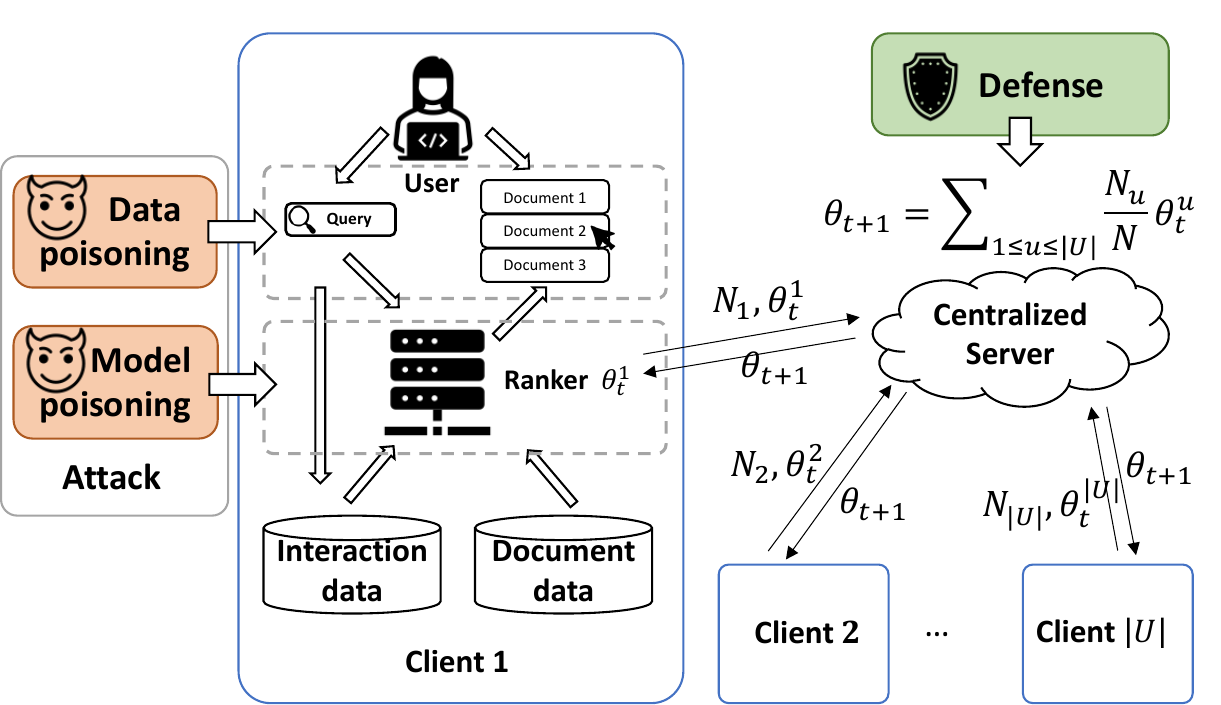}
	\vspace{-12pt}
	\caption{Overview of a FOLTR system with attack and defense modules (the arrows point to where these modules will be applied to).
	\label{fig:foltr}}
	\vspace{-12pt}
\end{figure}

In Online Learning to Rank (OLTR), all documents are stored in a server, and users' queries and interaction data (e.g., clicks) are also collected in the server. The ranker is then trained in a centralised and online manner.
However, this setting could potentially infringe on users' privacy as users may not want to share their queries and interactions. In addition, documents containing personal information, like in email search~\cite{kim2017understanding} or desktop search~\cite{cohen2008ranking}, may not be appropriate to surrender to a third party search service.
To address this issue, a new paradigm -- Federated Online Learning to Rank (FOLTR) -- has been explored~\cite{kharitonov2019federated,wang2021effective,wang2021federated,wang2022non}. 
In FOLTR (as in Figure~\ref{fig:foltr}), clients retain their data locally, train a local ranker, and then share the local model weights (or gradients) with the sever instead of the raw data. The server plays a very different role -- aggregating the received weights in an effective manner (e.g., via federated averaging~\cite{mcmahan2017communication}) and then broadcasting the obtained global ranker to the clients, which in turns use the global ranker to replace their local ranker.
The whole process is carried out iteratively.
Compared with conventional OLTR, FOLTR provides a mechanism to safeguard users' privacy. Also, the collaborative training makes the local rankers more effective than if they were trained separately with only the data of each single user. 



Existing FOLTR systems however are not necessarily secure: the federation mechanism provides malicious clients with opportunities for attacking the effectiveness of the global ranker. 
For example, malicious clients can send arbitrary weights to the server so that the convergence of the global ranker can be perturbed after aggregation. This kind of attack is termed as \textit{untargeted poisoning attack} and aim to compromise the integrity of the global model trained federatively~\cite{bagdasaryan2020backdoor}.
This issue is critical for federated learning systems, but it has not yet been studied for FOLTR. In this work, we initiate the investigation of poisoning attacks and corresponding defense methods in the context of FOLTR systems.

%
%
Outside of FOLTR systems, poisoning attacks on federated learning systems has been shown successful in compromising model integrity across several federated machine learning tasks~\cite{biggio2012poisoning, shafahi2018poison, bhagoji2019analyzing, lyu2020threats, yu2023untargeted}, including in natural language processing and recommender systems. 
To mitigate or remove the threat posed by poisoning attacks, defense strategies have been designed and optimised~\cite{blanchard2017machine, yin2018byzantine, guerraoui2018hidden}. 
Defense strategies typically act upon the aggregation rules used in the global model updating phase.
The vulnerability of existing FOLTR methods to these attacks and the effectiveness of the related defense mechanism is unknown. 
Previous work in FOLTR has shown that findings obtained with respect to federated learning in other areas of Machine Learning or Deep Learning do not directly translate to the online learning to rank context, and therefore the study of these techniques in the context of FOLTR is important. For example, \citet{wang2022non} have found that methods for dealing with non identical and independently distributed data in federated learning systems do not generalise to the context of FOLTR. 
%
Therefore, the performance of poisoning attacks and defense methods proposed in general domain cannot be guaranteed when applied to FOLTR: we address this limitation by adapting and investigating these methods to the setting of FOLTR and establish baselines for future studies.

In this paper, we complement the state-of-the-art FOLTR system with one untargeted attack module and one defense module (shown in Figure~\ref{fig:foltr}).
For the untargeted attack module, we implement a \textit{data poisoning} method that compromises the local training data to affect the trained model, and two \textit{model poisoning} methods that directly corrupt the local model updates. 
As for the defense module, we implement four Byzantine-robust aggregation rules to safeguard against such attacks. These defense mechanisms rely on statistical techniques to identify outliers among the received weights and subsequently exclude them during the aggregation process.
Through extensive empirical experiments, we 
(1) investigate the vulnerability of FOLTR systems to untargeted poisoning attacks, and show under which conditions poisoning attacks can represent a real threat to FOLTR systems; and
(2) demonstrate the effectiveness of defense strategies, and importantly reveal the presence of issues with defense strategies if applied to FOLTR systems for which an attack is not in place.



\section{Related Work}

\subsection{Federated OLTR}

Unlike traditional Learning to Rank (LTR), Online Learning to Rank (OLTR) optimizes rankers through implicit user feedback (e.g., clicks) to directly influence search engine result pages in real-time production. The earliest method, Dueling Bandit Gradient Descent (DBGD)~\cite{yue2009interactively}, uniformly samples variations of the ranking model and updates the ranker based on online interleaving evaluation. To mitigate the high variance and regret inherent in DBGD, subsequent methods have improved it through techniques like multiple interleaving~\cite{schuth2016multileave, oosterhuis2017balancing}, projected gradient~\cite{wang2019variance}, and counterfactual evaluation~\cite{zhuang2020counterfactual}. In contrast to DBGD-based approaches, Pairwise Differentiable Gradient Descent (PDGD)~\cite{oosterhuis2018differentiable} utilizes a Plackett-Luce model to sample the ranking list and estimates gradients from inferred pairwise preferences. This method has been found to exhibit greater resilience to noise and higher effectiveness in optimizing neural models.

OLTR methods have been thoroughly investigated in a centralized setting, where a central server possesses the data to be searched and gathers users' search interactions, such as queries and clicks. The training of the ranker also takes place on this server. However, this centralized paradigm is not well-suited for privacy-preserving requirement where each client may not wish to, or cannot, share the searchable data, queries and other interactions. This is the case, for example, of hospitals wanting to collaborate together to create powerful rankers to identify the cohort of patients for specific rare conditions (and as such, each hospital only holds limited data that would not be sufficient to train an effective ranker individually), but that by legislation they are forbidden to share the actual data.

To handle this issue, Federated Online Learning to Rank (FOLTR) methods have been proposed. These methods consider a decentralized machine learning scenario where data owners (clients) collaboratively train the model without sharing their data under the coordination of a central server. One such method is the Federated OLTR with Evolutionary Strategies (FOLtR-ES)~\cite{kharitonov2019federated}, which extends the OLTR optimization scenario to the Federated SGD~\cite{mcmahan2017communication} and utilizes Evolution Strategies as optimization method~\cite{salimans2017evolution}. While FOLtR-ES performs well on small-scale datasets under certain evaluation metrics, its effectiveness does not generalise to large-scale datasets and standard OLTR metrics~\cite{wang2021federated}. Because of this, we do not consider FOLtR-ES in our study. An alternative method is the FPDGD~\cite{wang2021effective}, which builds upon the state-of-the-art OLTR method, the Pairwise Differentiable Gradient Descent (PDGD)~\cite{oosterhuis2018differentiable}, and integrates it into the Federated Averaging (FedAvg) framework~\cite{mcmahan2017communication}. FPDGD exhibits effectiveness comparable to centralized OLTR methods, representing the current state-of-the-art FOLTR method. Thus, our empirical investigation of attack and defense methods on FOLTR systems relies on the FPDGD method, which is further described in Section~\ref{foltr}.

\vspace{-12pt}
\subsection{Poisoning Attacks on Federated Learning}

Poisoning attacks on federated learning systems aim to compromise the integrity of the system's global model. Poisoning attacks can be grouped according to the goals of the attack into two categories: untargeted poisoning attacks, and targeted poisoning attacks (also known as backdoor attacks).

Targeted poisoning attacks aim to manipulate a global model according to the attacker's objectives, such as misclassifying a group of data with certain features to a label chosen by the attacker, while maintaining normal model effectiveness under other conditions. This is accomplished through backdoor attacks~\cite{bagdasaryan2020backdoor, bhagoji2019analyzing}, which are designed to allow the targeted manipulations to transpire stealthily and without detection. 


In contrast, untargeted poisoning attacks (also known as Byzantine failures~\cite{lamport1982byzantine, blanchard2017machine, yin2018byzantine, fang2020local, shejwalkar2021manipulating}) aim to decrease the overall effectiveness of the global model indiscriminately for all users and data groups. Current untargeted poisoning methods can be divided into two categories: data poisoning and model poisoning. 
Label flipping~\cite{biggio2012poisoning} is a representative data poisoning method:  the labels of honest training data are changed without altering their features.
Model poisoning, on the other hand, directly affects the local model updates before they are sent to the centralized server. For example, Baruch et al.~\cite{baruch2019little} poisons the local model updates through the addition of noise computed from the variance between the before-attack model updates, while Fang et al.~\cite{fang2020local}'s attacks are optimized to undermine specific robust aggregation rules.


Among untargeted poisoning attacks on federated learning systems, model poisoning methods have been found to be the most successful~\cite{bagdasaryan2020backdoor}. In particular, data poisoning attacks have limited success when Byzantine-robust defense aggregation rules are in use~\cite{fang2020local}; we introduce these defense methods in Section~\ref{defense}.
Furthermore, most data poisoning attacks assume that the attacker has prior knowledge about the entire training dataset, which is often unrealistic in practice.

In this paper, we focus on untargeted poisoning attacks, delving into the effectiveness of both data poisoning and model poisoning methods. These attack methods are studied within the framework of a FOLTR system based on FPDGD, with and without the integration of defense countermeasures. 


\section{PRELIMINARIES}
\label{foltr}

%

\subsection{Online Learning to Rank (OLTR)}
In OLTR, the ranker is learned directly from user interactions (clicks in our study), rather than editorial labels.
In this context, each client performs searches on several queries during each local training phase. For each query $q$, the candidate documents set is $D_q$ and the local training data held by each client is $\{(x_i, c_i),  i=1...|D_q|\}_q$ with feature representation ($x_i$) and user's click signal ($c_i$) for each $(q, d_i)$-pair (where $d_i \in D_q$). The value of the click feedback $c_i$ is either 0 (unclicked) or 1 (clicked). In practice, the $click$ is dependent on the relevance degree of the candidate document $d_i$ to the query $q$, the rank position of $d_i$, and other noise or randomness factors.

\subsection{Federated Pairwise Differentiable Gradient Descent (FPDGD)}



We add our attacking and defense modules to the current state-of-the-art FOLTR system,
the Federated Pairwise Differentiable Gradient Descent (FPDGD)~\cite{wang2021effective}, which is outlined in Algorithm~\ref{alg:fpdgd}.
%
%
%
%
Within each iteration $t$, each client $u$ considers $N_u$ interactions and updates the local ranker using Pairwise Differentiable Gradient Descent (PDGD)~\cite{oosterhuis2018differentiable}. 
After the local update is finished, each client sends the trained weights $\theta^u_t$ to the server. 
The server then  leverages the widely-used Federated Averaging~\cite{mcmahan2017communication} to aggregate the local model updates. 
Afterwards, the new global weights $\theta_{t+1}$ are sent back to the clients as their new local rankers. We refer the reader to the original FPDGD paper for more details~\cite{wang2021effective}. 

%
%
%

\begin{algorithm}[t]
	\caption{FederatedAveraging PDGD. \\
		-  set of clients participating training: $U$, each client is indexed by $u$; \\
		- local interaction set: $B$, number of local interactions: $N_u$.}
	\label{alg:fpdgd}
	\begin{algorithmic}[1]
		\SUB{Server executes:}
		\STATE initialize $\theta_0$; scoring function: $f$; learning rate: $\eta$
		\FOR{each round $t = 1, 2, \dots$}
		\FOR{each client $u \in U$ \textbf{in parallel}}
		\STATE $\theta_{t}^u, N_u \leftarrow \text{ClientUpdate}(u, \theta_t)$
		\ENDFOR
		\STATE $\theta_{t+1} \leftarrow \sum_{u=1}^{|U|} \frac{N_u}{\sum_{u=1}^{|U|} N_u} \theta_{t}^u$
		\ENDFOR
	\end{algorithmic}
	\[\]
	\vspace{-22pt}
	\begin{algorithmic}[1]	
	\SUB{ClientUpdate($u, \theta$):} \ \ \ // \emph{Run on client $u$}
	\FOR{each local update $i$ from $1$ to $N_u$}
	\STATE $\theta \leftarrow \theta + \eta\nabla f_\theta$    \hfill \textit{\small //PDGD update with data from $B$} 
	\ENDFOR
	\STATE return ($\theta, N_u$) to server
	\end{algorithmic}	
\end{algorithm}

\section{Attacks to FOLTR Systems}
\subsection{Problem Definition and Threat Model}


\textbf{Attacker's capability:} Poisoning attacks can come from both members (insiders) and non-members (outsiders) of the FOLTR system. Insiders include both the central server and the clients, while outsiders include eavesdroppers on communication channels and users of the final  ranker  (this is similar to adversarial attacks during inference).
In this study, we focus on insider attacks by malicious participants in the FOLTR system since insider attacks are generally more effective than outsider attacks~\cite{lyu2020threats}. 
 
\noindent
We assume the attacker has control over $m$ collusive clients, which means that the training data and local model updates can be exchanged among the malicious clients. We restrict the percentage of collusive clients to less than 50\%: higher amounts would make it trivial to manipulate the global model.

\textbf{Attacker's background knowledge:} 
We assume that the attacker has only access to the compromised clients: the training data and local rankers of all remaining clients remain not accessible to the attacker.
Thus, the attacker has limited prior knowledge: the training data and the locally updated models from the poisoned clients, and the shared global model. 
The exception of having full prior knowledge\footnote{i.e. the attacker can also access information (training data, model updates) of non-poisoned clients.} will only be for the purpose of analysis and will be clarified in place.


\textbf{Problem Formulation:} Assume $n$ clients are involved in the FOLTR system. Among them, $m$ clients are malicious.
Without loss of generality, we assume the first $m$ participants are compromised. Be $\mathbf{w_i}$ the local model that the $i$-th client sends to the central server. The global ranking model is updated through aggregating all $\mathbf{w_i}$:

\begin{equation}
	\mathbf{w_g} = agg(\mathbf{w_1}, ..., \mathbf{w_m}, \mathbf{w_{m+1}}, ..., \mathbf{w_n})
	\label{eq:agg}
\end{equation}


\subsection{Data Poisoning}
\label{data-poison}

Data poisoning methods aim to corrupt the training data in order to degrade the model's effectiveness. This can be done by adding malicious instances or altering existing instances in an adversarial manner.




Our data poisoning attack to FOLTR is inspired by the label flipping strategy~\cite{biggio2012poisoning, tolpegin2020data}, in which the labels of honest training samples from one class are flipped to another class, while the features of the flipped samples are kept unchanged. 
In our case, we  want to change the label of irrelevant documents into "high-relevant" and vise versa, without any changes to the feature representation of the corresponding query-document pairs. 
%
To achieve so, the attacker needs to intentionally flip the feedback by clicking on irrelevant documents to bring arbitrary noise thus poison the training.

In our experiments, as no click data is available with the considered datasets, we follow the common practice from previous literature in OLTR and FOLTR~\cite{oosterhuis2018differentiable, wang2021effective, wang2022non} of simulating click behaviour based on the extensively-used \textit{Simplified Dynamic Bayesian Network} (SDBN) click model~\cite{chapelle2009dynamic}. This click model has been shown to produce reasonable predictions of real-world user click behaviour. Under SDBN, users examine a search engine result page (SERP) from top to bottom. Each document is inspected and clicked with click probability $P(click = 1|rel(d))$, conditioned on the actual relevance label $rel(d)$ of the document. After a document is clicked, the user decides to stop the search session with stopping probability $P(stop = 1|click=1, rel(d))$, or continue otherwise. Commonly, three instantiations of SDBN are considered: (1) a \textit{perfect} user examines every document and clicks on all relevant documents thus provides very reliable feedback, (2) a \textit{navigational} user searches for reasonably relevant documents with a higher probability to stop searching after one click, (3) an \textit{informational} user typically clicks on many documents without a specific information preference thus provides the noisiest click feedback. 

Inspired by the three instantiations, we manipulate one \textit{poison} instantiation to simulate malicious clicking behaviour. The click probability of \textit{poison} instantiation is the reverse version of the \textit{perfect} click behaviour: the highest probability of clicking is associated with the least relevance label. All stop probabilities in \textit{poison} instantiation are set to zero as we assume the attacker wants to poison as many clicks as possible. The values we adopt for the four instantiations of SDBN are reported in Table~\ref{tbl:sdbn}.


\begin{table}[t]
	\centering
	\caption[centre]{Instantiations of SDBN click model for simulating user behaviour in experiments. $rel(d)$ denotes the relevance label for document $d$. Note that in the MQ2007 dataset, only three-levels of relevance are used. We demonstrate the values for MQ2007 in bracket. \vspace{-14pt}}
	\begin{tabular}{p{1.4cm}<{\centering} p{1.3cm}<{\centering} p{1.1cm}<{\centering} p{1.2cm}<{\centering} p{0.75cm}<{\centering} p{0.85cm}<{\centering}}
		\hline
		& \multicolumn{5}{c}{$P(\mathit{click}=1\mid rel(d))$}  \\
		\cmidrule(r){2-6}
		\emph{rel(d)} & 0 & 1 & 2 & 3 & 4 \\
		\hline
		\emph{perfect} &  0.0 (0.0)&  0.2 (0.5)&  0.4 (1.0)&  0.8 (-)&  1.0 (-)\\
		\emph{navigational} & 0.05 (0.05)& 0.3 (0.5)&  0.5 (0.95)&  0.7 (-)&  0.95 (-)\\
		\emph{informational} &  0.4 (0.4)&  0.6 (0.7)&  0.7 (0.9)&  0.8 (-)&  0.9 (-)\\
		\emph{poison} &  1.0 (1.0)&  0.8 (0.5)&  0.4 (0.0)&  0.2 (-)&  0.0 (-)\\
		\hline\hline
		& \multicolumn{5}{c}{$P(\mathit{stop}=1\mid click=1,  rel(d))$} \\
		\cmidrule(r){2-6}
		\emph{rel(d)} & 0 & 1 & 2 & 3 & 4 \\
		\hline
		\emph{perfect}  & 0.0 (0.0)& 0.0 (0.0)&  0.0 (0.0)&  0.0 (-)&  0.0 (-)\\
		\emph{navigational} & 0.2 (0.2)&  0.3  (0.5)&  0.5 (0.9)&  0.7 (-)&  0.9 (-)\\
		\emph{informational} & 0.1 (0.1)&  0.2 (0.3)&  0.3 (0.5)&  0.4 (-)&  0.5 (-)\\
		\emph{poison}  & 0.0 (0.0)& 0.0 (0.0)&  0.0 (0.0)&  0.0 (-)&  0.0 (-)\\
		\hline
	\end{tabular}
	\label{tbl:sdbn}
	\vspace{-15pt}
\end{table}

\vspace{-6pt}
\subsection{Model Poisoning}
\label{model-poison}

Unlike data poisoning, model poisoning directly modifies the local model updates (through poisoning gradients or model parameter updates) before sending them to the server. Some literature shows that model poisoning is more effective than data poisoning~\cite{bagdasaryan2020backdoor, bhagoji2019analyzing} while it also requires sophisticated technical capabilities and high computational resources than solely poisoning data. In this section, we investigate two existing model poisoning methods.

\subsubsection{\textbf{Little Is Enough (LIE)}}

Baruch et al.~\cite{baruch2019little} find that if the variance between local updates is sufficiently high, the attacks can make use of this by adding small amounts of noise to the compromised local models and bypass the detection of defense methods. 
They provide a perturbation range in which the attackers can successfully poison the learning process. To conduct the attack, the adversaries first compute the average $\mu$ and standard deviation $\sigma$ of the before-attack benign local model updates of all collusive attackers ($\mathbf{w_1}, ..., \mathbf{w_m}$). A coefficient $z$ is used and computed based on the number of benign and malicious clients. Finally, the local model of attackers is manipulated as $\mathbf{w^m_i} = \mu - z\sigma$ for $i \in \{1, ..., m\}$ and sent to the central server who aggregates updates from all participants under certain rules in Equation~\ref{eq:agg}. Baruch et al.~\cite{baruch2019little} observe that, for image classification tasks, the small noises sufficiently compromise the global model while being sufficiently small to evade detection from defense strategies.

\vspace{-7pt}
\subsubsection{\textbf{Fang's Attack}}
\label{sec:fangattack}

Fang et al.~\cite{fang2020local} proposed an optimization-based model poisoning attack tailored to specific robust aggregation rules (Krum, Multi-Krum, Trimmed Mean and Median), as will be explained in Section~\ref{defense}. 

Fang's attack is conducted separately under two assumptions: (1) full knowledge, and (2) partial knowledge. Under full knowledge, the attacker has full access to local model updates of all benign clients. This is a strong and impractical assumption, and it is often not the case in real attacks on federated learning systems.
 In the partial knowledge scenario, the attacker only knows the local training data and models of the compromised clients.

In their attack to the robust aggregation rules Krum and Multi-Krum, the attacker computes the average $\mu$ of the benign updates in their possession, computes a perturbation $\mathbf{s}= - \text{sign}(\mu - \mathbf{w_g})$, and finally computes a malicious update as $\mathbf{w^m_i} = (\mathbf{w_g} + \lambda\cdot\mathbf{s})$ by solving for the coefficient $\lambda$, where $\mathbf{w_g}$ is the before-attack global model during each federated training step. Thus, under the full knowledge assumption, the average $\mu$ and perturbation signal $\mathbf{s}$ are computed based on all benign updates ($\mathbf{w_1}, ..., \mathbf{w_m}, \mathbf{w_{m+1}}, ..., \mathbf{w_n}$). For updates of the malicious clients ($\mathbf{w_1}, ..., \mathbf{w_m}$), the before-attack benign updates are leveraged. Under the partial knowledge scenario, only the before-attack benign updates ($\mathbf{w_1}, ..., \mathbf{w_m}$) are used to estimate the real values for average $\mu$ and the reversed deviation vector $\mathbf{s}$.

When attacking Trimmed Mean and Median, the goal is to craft the compromised local models based on the maximum $w_{max,j}$ or minimum $w_{min,j}$ benign parameters for each dimension $j$ of the local model (this is one of the key features used by Trimmed Mean and Median for defending). The choice of $w_{max,j}$ or $w_{min,j}$ depends on which one deviates the global model towards the inverse of its update direction without attacks. 
Similar to when attacking Krum, the reversed deviation vector $\mathbf{s}$ is computed with full knowledge or estimated under partial knowledge with only before-attack updates from all attackers, so as the estimation of $w_{max,j}$ and $w_{min,j}$. After getting the $j$-th value of vector $\mathbf{s}$, the $j$-th dimension of the compromised local model is randomly sampled from the range built based on $w_{max,j}$ (if $s_j = 1$) or $w_{min,j}$ (if $s_j = -1$).

\section{Defense for FOLTR Systems}
\label{defense}

%
%

The current state-of-the-art defense methods against untargeted poisoning attacks focus on enhancing the robustness of the aggregation rules (Equation~\ref{eq:agg}) used during the global update phase, to counteract attempts by malicious clients to corrupt the training.

Next, we describe four robust aggregation rules that have been shown effective in general federated learning, but have not been evaluated for FOLTR.

\subsection{Krum and Multi-Krum}

The intuition behind the Krum method for robust aggregation~\cite{blanchard2017machine} is that the malicious local model updates need to be far from the benign ones in order for the success of poisoning the global model. To evaluate how far a model update $\mathbf{w_i}$ is from the others, Krum computes the Euclidean distances between $\mathbf{w_i}$ and $\mathbf{w_j}$ for $i \neq j$. We denote $i \rightarrow j$ if $\mathbf{w_j}$ belongs to the set of $n-m-2$ closest local models of $\mathbf{w_i}$. Then the sum of $n-m-2$ shortest distances to $\mathbf{w_i}$ is computed and denoted as $s(i) = \sum_{i \rightarrow j}^{}Euc\_dist(\mathbf{w_i}, \mathbf{w_j})$. After computing the distance score $s(i)$ for all local updates, Krum selects the local model with the smallest $s(i)$ as the global model $w_g$:

\begin{equation}
	\begin{split}
	\mathbf{w_g} = \textit{Krum}(\mathbf{w_1}, ..., \mathbf{w_m}, \mathbf{w_{m+1}}, ..., \mathbf{w_n})
	= \mathop{\arg\min}_{\mathbf{w_i}}\ s(i)
	\end{split}
	\label{eq:krum}
\end{equation}

Multi-Krum is a variation of the Krum method. Multi-Krum, like Krum, calculates the distance score $s(i)$ for each $\mathbf{w_i}$. However, instead of choosing the local model with the lowest distance score as the global model (as Krum does), Multi-Krum selects the top $f$ local models with the lowest scores and computes the average of these $f$ models ($\mathbf{w'_i}$, where $i \in \{1, ..., f\}$) to be the global model.

\vspace{-7pt}
\begin{equation}
	\begin{split}
		\mathbf{w_g} = \textit{Multi-Krum}(\mathbf{w_1}, ..., \mathbf{w_m}, \mathbf{w_{m+1}}, ..., \mathbf{w_n})
		=\frac{1}{f} \sum_{i=1}^{f} \mathbf{w'_i}
	\end{split}
	\label{eq:multi-krum}
\end{equation}
\vspace{-7pt}

\noindent In our empirical investigation, we set the Multi-Krum parameter $f =  n-m$, as in previous work~\cite{blanchard2017machine}.

\subsection{Trimmed Mean and Median}

Assume that $w_{ij}$ is the $j$-th parameter of the $i$-th local model. For each $j$-th model parameter, the Trimmed Mean method~\cite{yin2018byzantine} aggregates them separately across all local models. After removing the $\beta$ largest and smallest among $w_{1j}, ..., w_{nj}$, the Trimmed Mean method computes the mean of the remaining $n-2\beta$ parameters as the $j$-th parameter of the global model. We denote $U_j = \{w_{1j}, ..., w_{(n-2\beta)j}\}$ as the subset of $\{w_{1j}, ..., w_{nj}\}$ obtained by removing the largest and smallest $\beta$ fraction of its elements. That is, the $j$-th parameter of the global model updated by Trimmed Mean is:
\vspace{-4pt}
\begin{equation}
	w_j = \textit{Trimmed Mean}(w_{1j}, ..., w_{nj}) = \frac{1}{n-2\beta} \sum_{w_{ij}\in U_j} w_{ij}
	\label{eq:trimmed-mean}
\end{equation}
\vspace{-6pt}

\noindent In our implementation, as in previous work on general federated learning~\cite{fang2020local, shejwalkar2021manipulating, shejwalkar2022back}, we set $\beta$ to be the number of compromised clients $m$.

The Median method, like the Trimmed Mean method, sorts the $j$-th parameter of $n$ local models. Instead of discarding the $\beta$ largest and smallest values (as in Trimmed Mean), the Median uses the median of $w_{1j}, ..., w_{nj}$ as the $j$-th parameter of the global model:

\begin{equation}
	w_j = \textit{Median}(w_{1j}, ..., w_{nj})
	\label{eq:median}
\end{equation}

\noindent In case $n$ is an even number, the median is calculated as the average of the middle two values.



\section{Experimental setup}

We next describe our experimental setup to evaluate the considered attack and defense mechanism in the context of a FOLTR system.


\textbf{Datasets.}
Our experiments are performed on four commonly-used LTR datasets: MQ2007~\cite{qin2013introducing}, MSLR-WEB10k~\cite{qin2013introducing}, Yahoo~\cite{chapelle2011yahoo}, and Istella-S~\cite{lucchese2016post}. Each dataset consists of a set of queries and the corresponding pre-selected candidate documents for each query. Each query-document pair is represented by a multi-dimensional feature vector, and have a corresponding annotated relevance label.
Among the selected four datasets, MQ2007~\cite{qin2013introducing} is the smallest with 1,700 queries, 46-dimensional feature vectors, and 3-level relevance assessments (from \textit{not relevant} (0) to \textit{very relevant} (2)). The other three datasets are larger, more recent, and provided by commercial search engines.  MSLR-WEB10k has 10,000 queries and each query is associated with 125 documents on average, each represented with 136 features. Yahoo has 29,900 queries and each query-document pair has 700 features. Istella-S is the largest, with 33,018 queries, 220 features, and an average of 103 documents per query. These three commercial datasets are all annotated for relevance on a five-grade-scale: from \textit{not relevant} (0) to \textit{perfectly relevant} (4).

\textbf{Federated setup.}
We consider 10 participants ($n=10$) in our experiments, among which $m$ clients are attackers. 
This setup is representative of a cross-silo FOLTR system, typical of a federation of a few institutions or organisations, e.g. hospitals creating a ranker for cohort identification from electronic health records~\cite{hersh2020information}. In this paper we will not consider the setup of a cross-device FOLTR system, where many clients are involved in the federation: this is representative of a web-scale federation. 

We assume that the malicious clients can collude with each other to exchange their local data and model updates to enhance the impact of attacks. In the federated setting, each client holds a copy of the current ranker and updates the local ranker through issuing $N_u=5$ queries along with the respective interactions. The attackers can only compromise the local updating phase through poisoning the training data or model updates of the controlled malicious clients. After the local updating finishes, the central server will receive the updated ranker from each client and aggregate all local messages to update the global ranker. In our experiments, we consider the following aggregation rules: (1) FedAvg, (2) other robust aggregation rules introduced in Section~\ref{defense}. Unless otherwise specified, we train the global ranker through $T=10,000$ global updating times.


\textbf{User simulations.}
We follow the standard setup for user simulations in OLTR~~\cite{oosterhuis2018differentiable, zhuang2020counterfactual, wang2021effective, wang2022non}. We randomly sample from the set of queries in the static dataset to determine the query that the user issues each time. After that, the pre-selected documents for the query are ranked by the current local ranking model to generate a ranking result. For every query, we limit the SERP to 10 documents. User interactions (clicks) on the displayed ranking list are simulated through the SDBN click models introduced in Sec~\ref{data-poison}. For the user simulation in model poisoning, we simulate three types of users using the three click instantiations: perfect, navigational, and informational. We experiment on the three types of users separately in order to show the impact of attacking on different types of users. For data poisoning, we simulate the poisoned click based on the \textit{poison} click combining with benign users on the aforementioned three types of click models separately to show the impact of our data poisoning strategies on different types of benign clicks. 

\textbf{Ranking models.}
We experiment on a linear and neural model as the ranking model when training with FPDGD. For the linear model, we set the learning rate $\eta = 0.1$ and zero initialization was used. As in the original PDGD and FPDGD studies~\cite{oosterhuis2018differentiable,wang2021effective}, the neural ranker is optimized using a single hidden-layer neural network with 64 hidden nodes, along with $\eta = 0.1$.

\textbf{Evaluation.}
We evaluate the attack methods by comparing the gap in offline performance obtained when a specific attack is performed and when no attack is performed. The higher the performance degradation, the more effective the attack.

As we limit each SERP to 10 documents, we use $nDCG@10$ for offline evaluation. The offline performance is measured through averaging the $nDCG$ scores of the global ranker over the queries in the held-out test dataset with the actual relevance label. We record the offline $nDCG@10$ score of the global ranker during each federated training update. 


\begin{figure*}[t]
	\centering
	\subfigure[\textbf{MSLR-WEB10k (Krum)}] {\label{fig:mslr-data-krum} \includegraphics[width=1\textwidth]{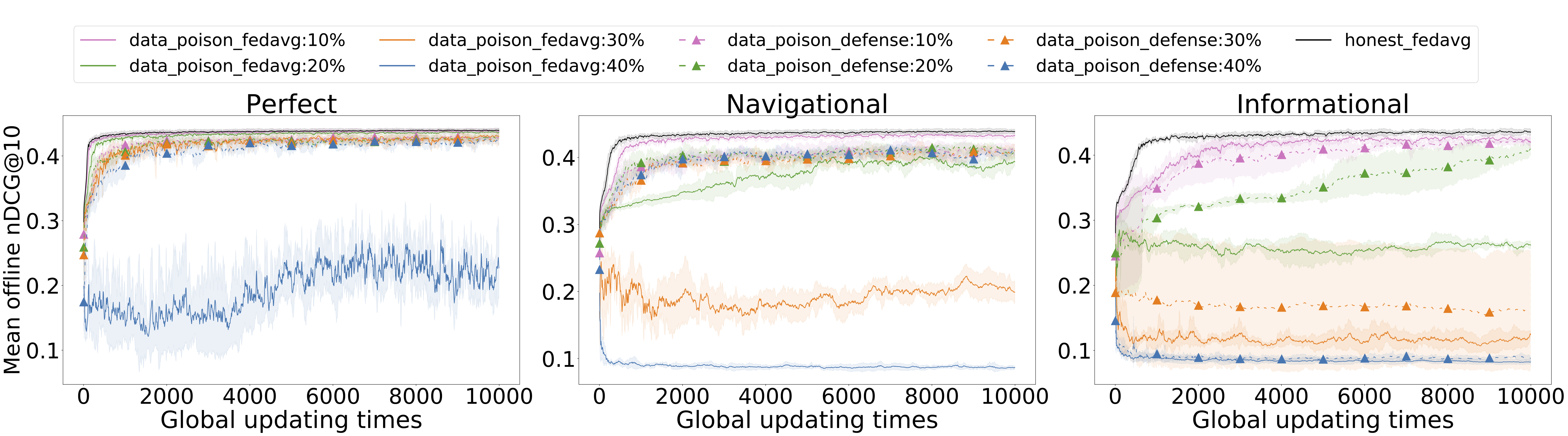}} 
	\subfigure[\textbf{MSLR-WEB10k (Multi-Krum)}] {\label{fig:mslr-data-multikrum} \includegraphics[width=1\textwidth]{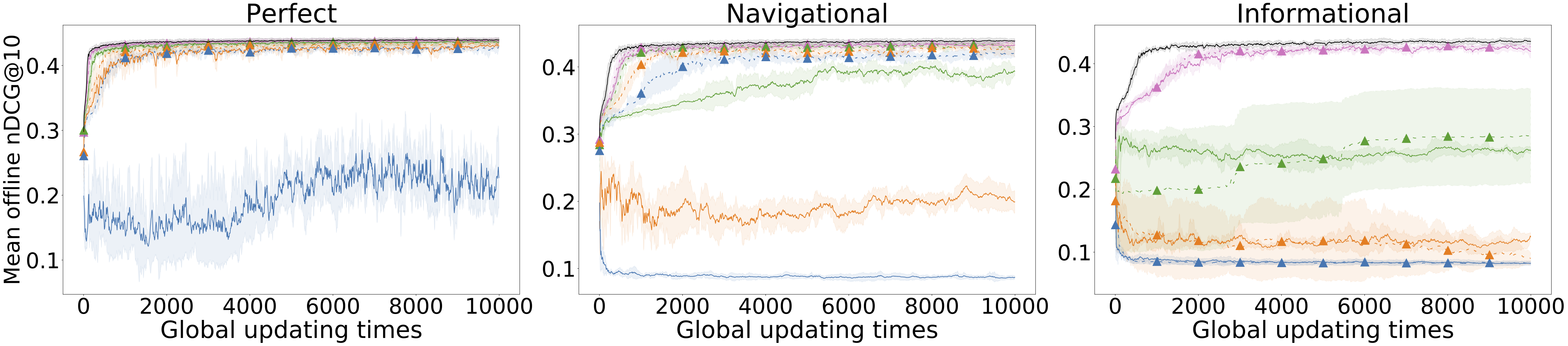}} 
	\subfigure[\textbf{MSLR-WEB10k (Trimmed Mean)}] {\label{fig:mslr-data-trmean} \includegraphics[width=1\textwidth]{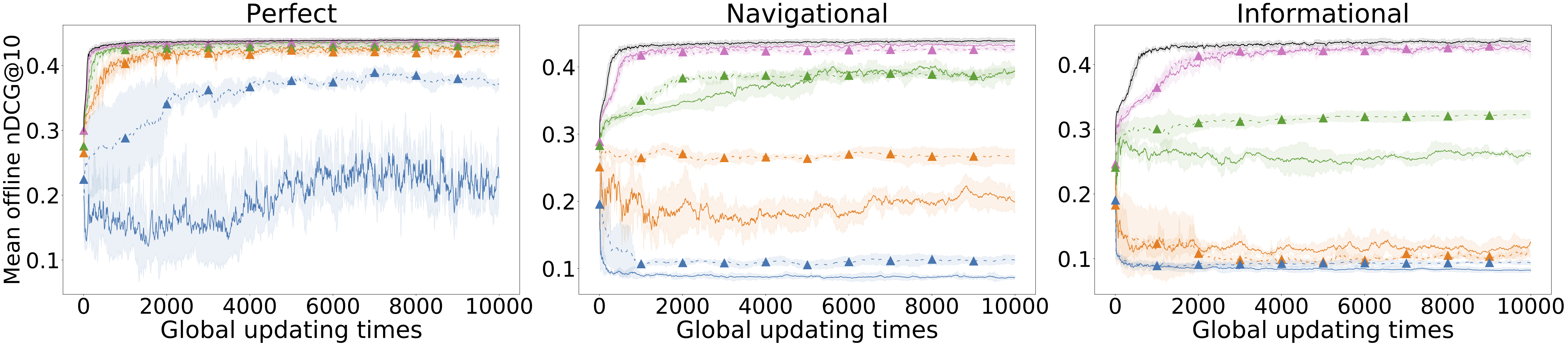}} 
	\subfigure[\textbf{MSLR-WEB10k (Median)}] {\label{fig:mslr-data-median} \includegraphics[width=1\textwidth]{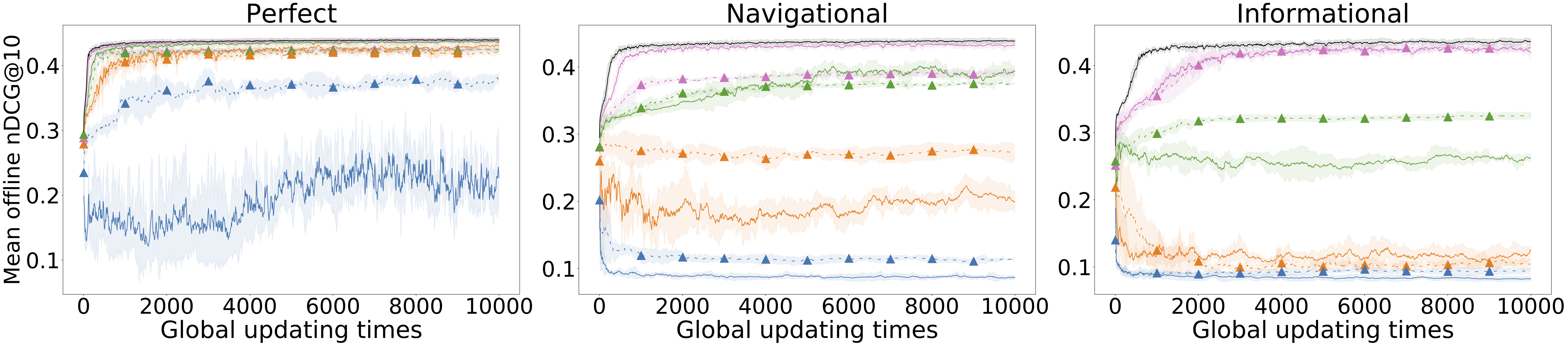}} 
	\vspace{-12pt}
	\caption{Offline performance (nDCG@10) for MSLR-WEB10k under data poisoning attack and defense strategies, simulated with three benign instantiations of SDBN click model and different percentage of attackers equaling to $\{10\%, 20\%, 30\%, 40\%\}$; results averaged across all dataset splits and experimental runs.
		\label{fig:data-defense}}
\end{figure*}

\section{Results for Data Poisoning}
We perform data poisoning attack and four defense methods across different settings of user behaviours (i.e. click models) and number of attackers ($\{10\%, 20\%, 30\%, 40\%\}$). 
Results on MSLR10k with a linear ranker are shown as solid lines in Figure~\ref{fig:data-defense} -- results for other datasets are similar and omitted for space constraints.

\subsection{Attacks}

%
%
%
%
%

In the plots of Figure~\ref{fig:data-defense}, the solid lines represent the results of data poisoning when no defense method is deployed. Among them, the black line represents no attacking situation ("honest" baseline).
%
We can observe that
the effect of data poisoning depends on the settings of user behaviors (i.e. click models) and the number of attackers.

\textbf{Effect of number of attackers.}
By comparing the solid curves in each plot of Figure~\ref{fig:data-defense}, we can observe that the overall performance of the FOLTR system decreases as the number of attackers increases, compared to the ``honest'' baseline.  Thus, the higher the number of attackers, the more degradation on the FOLTR system is experienced.

\textbf{Ease of attack under different user behaviours.}
By comparing the plots within each row, we see the effect of data poisoning is different under different user behaviours.
%
In the navigational and informational settings, attacks carried by as little as 20\% of clients can significantly affect the system. However, to successfully attack the perfect click, a higher number of malicious clients is needed. 
Across all datasets, the informational click model is the most affected by attacks, while the perfect click model only experiences considerable losses when a large number of clients has been compromised.


\textbf{Neural ranker vs. linear ranker.}
The findings from results for the neural ranker under data poisoning attack are similar to those for the linear ranker -- and this pattern is valid across all remaining experiments we report. Therefore, we only report experiments using the linear ranker due to limited space. 



\subsection{Defense}

%

Next, we demonstrate  the effectiveness of our four defense mechanisms against data poisoning attack.
The results on MSLR10k are shown by the dashed curves in Figure~\ref{fig:data-defense}. Each row corresponds to one defense method.

\textbf{Krum.} 
Overall, Krum performs well across all datasets and for all three types of click models once the percentage of malicious clients reaches 20\% or higher, with the exception of MQ2007. However, Krum does not work when defending against 10\% of clients, except for Istella-S. 
The accuracy drop from deploying Krum (as shown in Section~\ref{benign-defense}) outweighs its effectiveness in defense, especially when there is a relatively small impact on the effectiveness of the model, as is in the case when 10\% of the clients are malicious. 
Additionally, Krum does not show any improvement in defending certain scenarios under the informational click model, such as for MQ2007 under all percentages of malicious clients, and for MSLR10k when 40\% of clients are malicious.


\textbf{Multi-Krum.} 
The results obtained for Multi-Krum show similar effectiveness on the perfect click model as Krum. It is important to note that the perfect click model is the hardest to attack among the three types of click models considered. Multi-Krum provides slightly better defense performance on navigational clicks compared to Krum, especially when there are fewer attackers (30\% or less). However, for the informational click model, Multi-Krum does not perform as well as Krum. This is because the variance of the local model updates is relatively higher in the noisier informational click model. After averaging the selected local models, the advantage of Multi-Krum is reduced, especially when there are more than 30\% malicious clients.


\textbf{Trimmed Mean.}
Across all experiments, Trimmed Mean does not perform well on the noisiest click model (informational) when there are more than 30\% malicious clients involved. When the malicious clients are 20\% or 30\%, Trimmed Mean provides lower performance gains compared to Krum, but it performs similarly to Krum when only 10\% of the clients are malicious. 


\textbf{Median.}
Like Trimmed Mean, Median does not provide improved performance on the noisy informational click model when 30\% or 40\% of clients are malicious. Similarly, and like other robust aggregation rules, Median does not show significant improvements when only 10\% of  clients are malicious. In fact, the Median's performance even decreases on the navigational click model for MSLR10k with 10\% of malicious clients. When the malicious clients are 20\% and 30\% of all clients in the federation, the performance gain provided by Median is similar to that of Trimmed Mean.

\textbf{Summary.} Overall, Krum and Multi-Krum work better than Trimmed Mean or Median when defending against data poisoning attacks, with the exception that Trimmed Mean and Median perform better on the smaller MQ2007 dataset.

\begin{figure*}[t]
	\centering
	\subfigure[\textbf{MSLR-WEB10k (LIE)}] {\label{fig:mslr-lie} \includegraphics[width=1\textwidth]{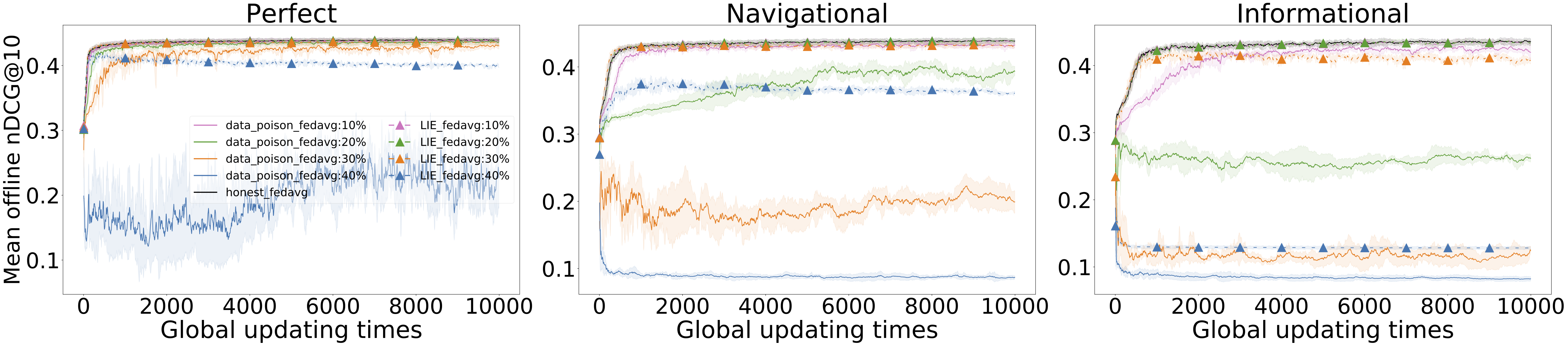}} 
	\subfigure[\textbf{MSLR-WEB10k (Fang's attack on Krum)}] {\label{fig:mslr-fang-krum} \includegraphics[width=1\textwidth]{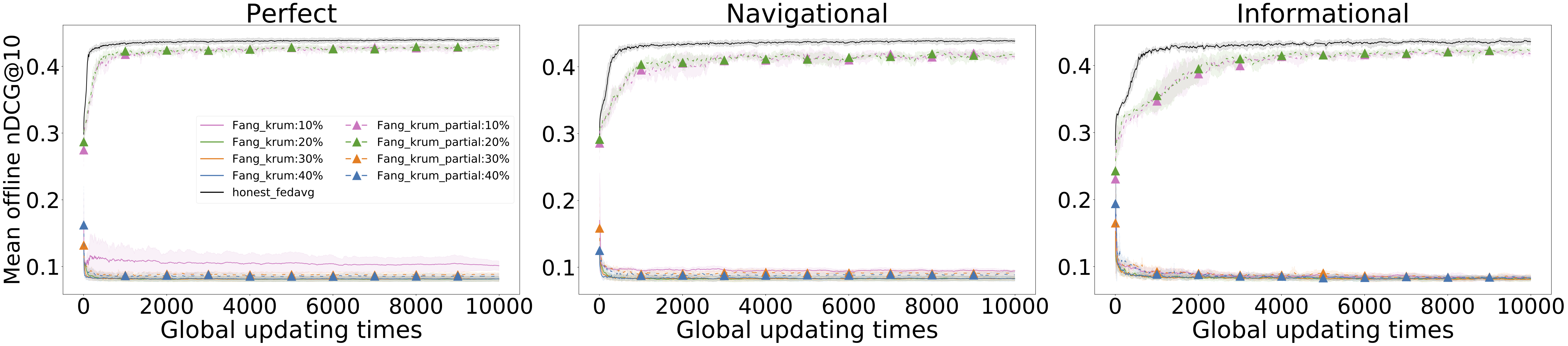}} 
	\subfigure[\textbf{MSLR-WEB10k (Fang's attack on Krum - full knowledge )}] {\label{fig:mslr-fang-krum-data} \includegraphics[width=1\textwidth]{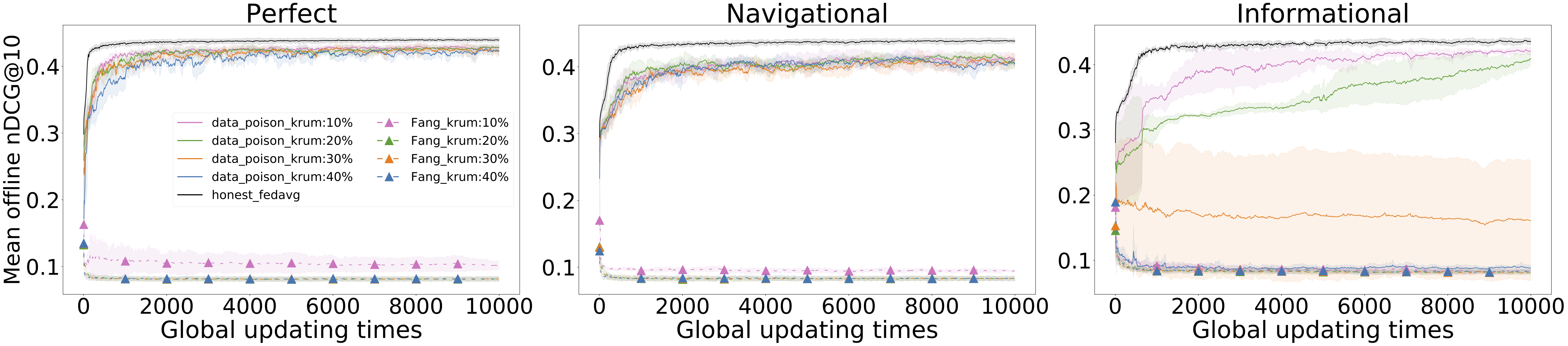}} 
	\vspace{-12pt}
	\caption{Offline performance (nDCG@10) under model poisoning attacks, simulated with three benign instantiations of SDBN click model under different percentage of attackers equaling to $\{10\%, 20\%, 30\%, 40\%\}$; results averaged across all dataset splits and experimental runs.
	\label{fig:model-attack}}
    \vspace{-12pt}
\end{figure*}

\section{Results for Model Poisoning}

We implement the model poisoning strategies specified in Section~\ref{model-poison} and report their results, specifically comparing their poisoning effectiveness with that of data poisoning methods.

\vspace{-8pt}
\subsection{Little Is Enough (LIE)}

The experimental results obtained for LIE are partially shown in Figure~\ref{fig:mslr-lie}, along with a comparison with data poisoning. 

\textbf{Ineffectiveness of LIE.}
The results indicate that LIE is less effective in attacking the performance of the global model compared to data poisoning, with one exception for the perfect click model on the Yahoo dataset when 40\% of the clients are malicious. This shows that adding random noise to compromise the local models is less effective for attacking the global ranker performance than compromising the click signals directly. Because of the poor attacking effectiveness of LIE, we do not investigate how it performs when defense strategies are put in place.


\subsection{Fang's Attack}
In our experiments, we implement Fang's attacks on four robust aggregation rules, with each attacking strategy tailoring specific defense strategies except that the same attack method is shared for Trimmed Mean and Median. 

\textbf{Full knowledge vs. partial knowledge.}
First, we compare the attacking performance under both full knowledge and partial knowledge assumptions. 
According to previous findings in general federated learning~\cite{fang2020local}, attacking with full knowledge performs consistently better than with partial knowledge as the tailored attack can be optimised with auxiliary information about benign clients. 
From our results (results on MSLR10k under Krum are shown in Figure~\ref{fig:mslr-fang-krum}), we observe that full knowledge performs better with fewer malicious clients (10\% and 20\%), but the gap in effectiveness obtained between full and partial knowledge decreases as the number of malicious clients increases (30\% and 40\%), thus leading to differences compared to the general results in federated learning. This is because with more malicious clients, partial knowledge (knowledge of before-attack local model updates for compromised clients) provides enough information to effectively poison the global model while avoiding detection by robust defense strategies.


\textbf{Fang's Attack vs. data poisoning.}
Next, we compare Fang's attack under the full knowledge assumption against the data poisoning method under the same robust-aggregation rule (results on MSLR10k under Krum are shown in Figure~\ref{fig:mslr-fang-krum-data}). We find that Fang's attack can successfully poison FOLTR and mitigate the impact of defense methods compared to data poisoning. This finding aligns with the original results from~\citet{fang2020local}.

\begin{figure*}[t]
	\centering
	\subfigure[\textbf{MSLR-WEB10k (benign clients)}] {\label{fig:mslr-benign-defense} \includegraphics[width=0.95\textwidth]{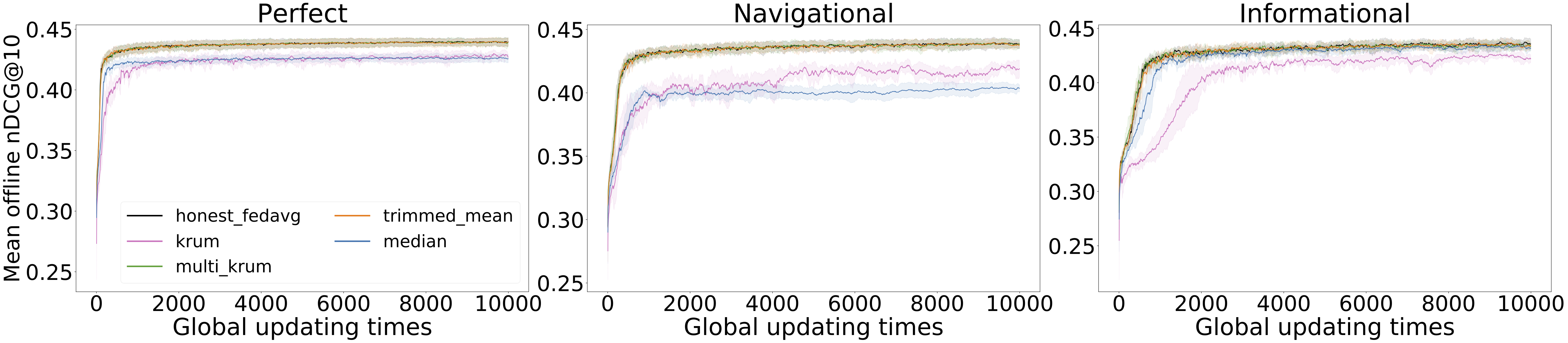}} 
	\vspace{-12pt}
	\caption{Offline performance (nDCG@10) of FOLTR system when no attack is present but defense strategies are deployed; results averaged across all dataset splits and experimental runs.
		\label{fig:benign-defense}}
	\vspace{-12pt}
\end{figure*}

\vspace{-8pt}
\section{Impact of Defense under No-Attack}
\label{benign-defense}

Robust aggregation rules exhibit improvements in defending against poisoning attacks under some circumstances. But in real-world settings, the administrator of the FOLTR system has no knowledge of whether an attack is taking place. Thus, if the system administrator wishes to ensure protection against attacks, they may be required to deploy defense strategies irrespective of an attack ever taking place, or not. However, is there a price to pay, in terms of search effectiveness, if a defense strategy is deployed on a FOLTR system that is not exposed to an attack? We investigate this next, by comparing the effectiveness of a FOLTR system with no malicious clients and with different defense strategies implemented against the effectiveness of the same system with no defense.

The experimental results on MSLR10k reported in Figure~\ref{fig:benign-defense} show that using Krum and Median leads to a decrease in performance compared to the FedAvg baseline when no attacks are present. Results for other datasets are similar and are omitted for space reasons. This finding has also been reported before in general federated learning literature~\cite{xia2019faba, cao2021fltrust, xu2022byzantine}, especially when each client's local training data is non independent and identically distributed (non-IID). This is because those Byzantine-robust FL methods exclude some local model updates when aggregating them as the global model update~\cite{cao2021fltrust, xia2019faba}. This decrease raises questions about the use of these methods in FOLTR systems when no malicious client is present -- and it suggests that if reliable methods for attack detection were available, then defense mechanisms may better be deployed only once the attack takes place. 



\vspace{-6pt}
\section{Summary of Key Findings}

Based on the presented empirical results above, we identify the following key findings:

\begin{itemize}
	\item In general, the perfect click type is more difficult to attack compared to the other two click models, whether it be data or model poisoning methods, except in specific instances when employing Fang's attack under the full knowledge assumption. To successfully attack a FOLTR system when perfect click feedback is present, a larger number of attackers is required due to the relatively low variance between local updates. As a result, more clients must be compromised to inject noise, otherwise the attack is more likely to be detected by robust aggregation rules.

\item Among all attacking strategies studied in this paper, Fang's attack with full knowledge emerged as the most successful in diminishing the performance of the global model, though some exceptions were observed in the noisy informational click scenario. When there were more malicious clients (i.e. 30\% or 40\% of the total clients), Fang's attack with partial knowledge is just as effective as with full knowledge. This indicates that model poisoning is more effective than data poisoning. Furthermore, when defense measures were implemented, Fang's attack demonstrated greater success against Krum and Multi-Krum aggregation rules in comparison to Trimmed Mean and Median.

	
	
\item It is essential to highlight that although Krum has proven effective in countering data poisoning and Trimmed Mean in defending against Fang's attack, deploying these two aggregation rules should be exercised with caution as they result in an overall decrease in search performance if the system is not exposed to attacks. Thus, the selection of these defense mechanisms should be carefully considered, taking into account the specific context and risk of potential attacks to strike the right balance between security and search effectiveness.

\end{itemize}
\vspace{-10pt}
\section{Conclusion}
In this paper we explore attacks and defense mechanisms for federated online learning to rank (FOLTR) systems, focusing on the potential degradation of ranking performance caused by untargedted poisoning attacks. We investigate both data and model poisoning strategies and evaluate the effectiveness of various state-of-the-art robust aggregation rules for federated learning in countering these attacks.
Our findings indicate that sophisticated model poisoning strategies outperform data poisoning methods, even when defense mechanisms are in place.
We also reveal that deploying defense mechanisms without an ongoing attack can lead to ranker performance degradation. This finding recommends care in the deployment of such mechanisms and suggests that future research should explore defense strategies that do not deteriorate FOLTR ranker performance if no attack is underway.

This is the first study that systematically analyses the threats brought by untargeted poisoning attacks and demonstrates the effectiveness (and associated drawbacks) of existing defense methods on mitigating the impact of malicious adversaries under federated online learning to rank system.

Due to space limitations, we could not include all experiment results in the paper. The complete results, along with code and settings are available at \url{https://github.com/ielab/foltr-attacks}.

\begin{acks}
Shuyi Wang is the recipient of a Google PhD Fellowship. This research is partially funded by Beijing Baidu Netcom Technology Co, Ltd, for the project "Federated Online Learning of Neural Rankers", under funding schema 2022 CCF-Baidu Pinecone.
\end{acks}

\bibliographystyle{ACM-Reference-Format}
\bibliography{ictir2023-foltr-poisoning}

%
%

\end{document}